\documentclass[11pt]{article}

\usepackage{algorithm}

\usepackage{amsmath,amssymb,amsfonts}
\usepackage{algorithmic}
\usepackage{graphicx}
\usepackage{textcomp}
\usepackage{xcolor}
\usepackage{enumitem}
\usepackage{hyperref}
\usepackage{lineno}
\usepackage[numbers]{natbib}
\usepackage{listings}
\usepackage{tikz}
\usepackage{booktabs}
\usepackage{float}  
\usepackage{amssymb}
\usepackage{pifont}

\usepackage{float}

\begin{document}

\title{Gradientsys: A Multi-Agent LLM Scheduler with ReAct Orchestration}

\author{%
  Xinyuan\,Song \\            
  Emory University \\          
  \texttt{xsong69@emory.edu}    
  \and
  Zeyu\,Wang \\
  Gradient Network \\
  \texttt{zeyuwang@ucla.edu}
  \and
  Siyi\,Wu \\
  University of Texas at Arlington \\
  \texttt{sxw8121@mavs.uta.edu}
  \and
  Tianyu\,Shi \\
  University of Toronto \\
  \texttt{ty.shi@mail.utoronto.ca}
  \and
  Lynn\,Ai \\
  Gradient Network \\
  \texttt{lynn@gradient.network}
}
\date{}          
\maketitle

\begin{abstract}
We present \textbf{Gradientsys}, a next-generation multi-agent scheduling framework that coordinates diverse specialized AI agents using a typed Model-Context Protocol (MCP) and a ReAct-based dynamic planning loop. At its core, Gradientsys employs an LLM-powered scheduler for intelligent one-to-many task dispatch, enabling parallel execution of heterogeneous agents such as PDF parsers, web search modules, GUI controllers, and web builders. The framework supports hybrid synchronous/asynchronous execution, respects agent capacity constraints, and incorporates a robust retry-and-replan mechanism to handle failures gracefully. To promote transparency and trust, Gradientsys includes an observability layer streaming real-time agent activity and intermediate reasoning via Server-Sent Events (SSE). We offer an architectural overview and evaluate Gradientsys against existing frameworks in terms of extensibility, scheduling topology, tool reusability, parallelism, and observability. Experiments on the GAIA general-assistant benchmark show that Gradientsys achieves higher task success rates with reduced latency and lower API costs compared to a MinionS-style baseline, demonstrating the strength of its LLM-driven multi-agent orchestration.
\end{abstract}

\section{Introduction}
\vspace{-0.5em}
Large Language Models (LLMs) augmented with tool use have enabled AI agents that can perform complex, multi-step tasks beyond basic question-answering~\cite{russell2021artificial}. By combining reasoning with actions (e.g. using tools or APIs), these agents can retrieve information, process documents, and interact with external environments to solve real-world problems. Recent benchmarks like GAIA target~\cite{gaia2025} such General AI Assistants, requiring abilities like multi-hop reasoning, web browsing, and tool-use in real scenarios. However, existing agent frameworks face limitations in scheduling and orchestration. Early systems like AutoGPT~\cite{richards2023autogpt} run a single agent loop sequentially, lacking true parallelism or modular reuse of specialized sub-agents. Other frameworks (e.g. Manas~\cite{koley2024manas}, TaskWeaver~\cite{taskweaver2024}) allow multiple agents or plugins, but often rely on static workflows or code-generation, and may not fully exploit dynamic LLM reasoning for on-the-fly planning.

We present \textbf{Gradientsys}, a multi-agent scheduling system that addresses these gaps by leveraging an LLM planner using the ReAct paradigm~\cite{yao2023react} to dynamically orchestrate a Gradientsys of tools. Gradientsys's core is an LLM-based scheduler that maintains a registry of available agents (tools) described in a standardized way via the Model Context Protocol (MCP)~\cite{hou2025mcp,anthropic2024mcp}. At runtime, the scheduler reasons about a given task, selects appropriate tools, and can dispatch multiple tool calls in parallel (one-to-many) when beneficial. It manages each agent's capacity (to avoid overload) and handles failures by retrying or replanning alternate strategies. To the user, Gradientsys provides improved observability: as the LLM planner reasons and agents perform work, the system streams intermediate steps and partial results live via Server-Sent Events (SSE)~\cite{whatwg_sse}, increasing transparency and debuggability.

\begin{figure*}[!ht]
\centering
\includegraphics[width=0.9\textwidth]{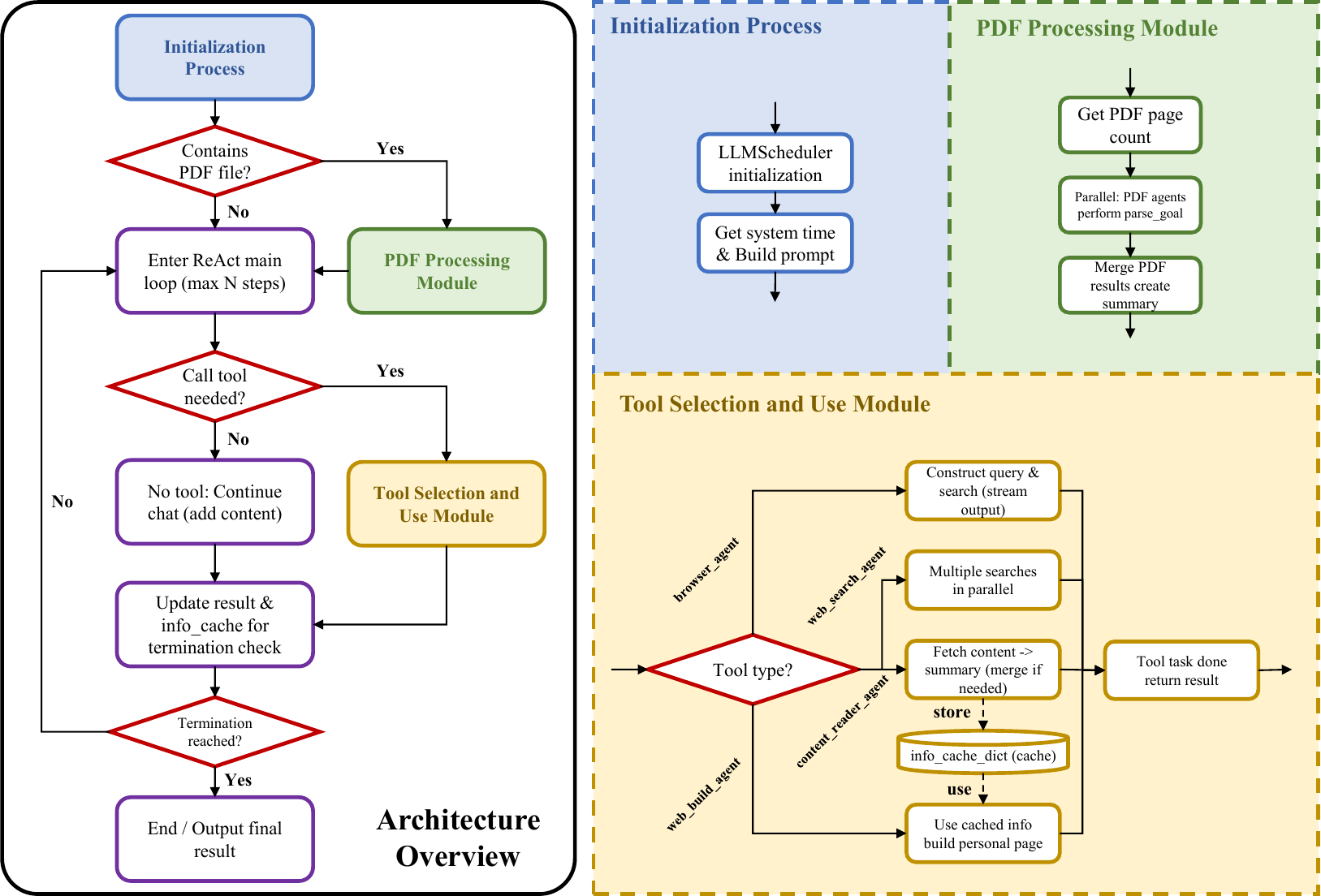}
\caption{Gradientsys architecture overview. The LLM Scheduler (center) employs a ReAct-based planner to coordinate with a registry of registered tools. It supports concurrent dispatch of actions to multiple agents (e.g., Agent A, Agent B), aggregates results via a summarization agent, and returns the final output. Dashed lines denote registry queries, while dotted lines represent observability event streams.}
\label{fig:architecture}
\end{figure*}

Gradientsys's design brings several key contributions:
\begin{itemize}[left = 0em]
\vspace{-0.5em}
\item \textbf{Dynamic LLM-Driven Planning:} We integrate the ReAct framework into the scheduler, allowing the LLM to generate interleaved reasoning traces and tool actions for adaptive task decomposition and real-time exception handling.
\vspace{-0.5em}
\item \textbf{Parallel Multi-Agent Orchestration:} The scheduler supports concurrent invocation of multiple agents when tasks can be parallelized (e.g., processing document chunks), unlike prior sequential frameworks. A typed MCP interface ensures seamless integration of heterogeneous tools with minimal custom code.
\vspace{-0.5em}
\item \textbf{Tool Registry and Extensibility:} Gradientsys provides a registry where tools register via a URL, a textual capability description, maximum parallelism, and optional cost metadata. This enables runtime hot-plugging of tools, with the LLM planner discovering and using them automatically. The standardized MCP-inspired interface improves reusability across tasks and LLM backends.
\vspace{-0.5em}
\item \textbf{Hybrid Sync/Async Execution:} The system supports a flexible mix of synchronous and asynchronous execution. The LLM planner can issue blocking calls for immediate results or launch asynchronous tasks while continuing reasoning. A threadpool-based dispatcher manages parallel agent calls while respecting each agent’s capacity.
\vspace{-0.5em}
\item \textbf{Observability and Control:} Planning traces and agent actions are streamed via SSE, enhancing transparency, debugging, and trust, and enabling real-time intervention if needed. The system is privacy-aware, sending each tool only the minimal necessary context to protect sensitive data.
\vspace{-0.5em}
\item \textbf{Improved Performance on GAIA Benchmark:} Gradientsys achieves higher accuracy and reduced latency and cost on the GAIA general-assistant benchmark compared to a MinionS-style baseline, demonstrating the effectiveness of its scheduling.
\end{itemize}

\vspace{-0.5em}
\section{System Architecture Overview}
\vspace{-0.5em}
\textbf{Gradientsys} is built around a centralized LLM Scheduler that orchestrates a pool of modular, specialized tool agents. Figure~\ref{fig:architecture} illustrates the high-level workflow. Upon receiving a user query (optionally with documents), the Scheduler initiates an iterative ReAct reasoning loop. It first queries the Tool Registry (Step 1) to discover available agents and their metadata. Based on this, it selects and dispatches appropriate tools, invoking multiple agents in parallel when subtasks are independent (Step 2).

Each agent (e.g., PDF parsers, web search modules, GUI controllers) performs its subtask and returns results asynchronously (Step 3). The Scheduler aggregates outputs, optionally delegating fusion to an Aggregator or Summarizer agent (Step 4), and returns the final result to the user (Step 5). Throughout, an Observability Module streams live updates—reasoning traces, tool invocations, intermediate results—via Server-Sent Events (SSE), enabling real-time monitoring and feedback.

A key strength of Gradientsys’s architecture is its integration with the Model-Context Protocol (MCP), a typed HTTP interface standard for connecting LLMs to tools. Each agent conforms to an MCP-compatible schema, enabling uniform, composable tool invocation.

The Tool Registry forms the system’s metadata backbone. Each tool is described by an endpoint, brief description, and optionally a type signature specifying I/O. Metadata includes \texttt{max\_parallel} (max concurrent calls) and \texttt{cost\_per\_1k tokens} (e.g., API cost). The scheduler uses this for planning and load management, preventing overload. Tools can be registered or removed at runtime, with the LLM context updated accordingly.

Gradientsys uses a hybrid sync/async dispatcher. When multiple tool actions are issued in a ReAct iteration (e.g., “Call AgentA with X and AgentB with Y”), they run concurrently via thread pools or coroutines. If a tool exceeds its concurrency limit, calls are queued or rerouted. Sync tasks block reasoning; async tasks allow overlap, reducing latency.

Central to orchestration is the ReAct loop, where the LLM alternates between thoughts (natural language) and actions (structured tool calls). Actions are parsed, executed, and results fed back into context. A scratchpad stores reasoning steps, allowing the LLM to track progress, avoid redundancy, and handle failures.

Observability is a core concern. At each planning and execution step—e.g., “Invoking Tool~$T$ with input~$X$,” “Tool~$T$ returned output~$Y$”—structured events are streamed via SSE to clients. This real-time transparency improves debugging and trust. Unlike frameworks like AutoGPT, which emit static logs post hoc, Gradientsys allows users to observe and intervene during execution.

\section{System Features and Implementation Details}
Gradientsys’s system-level features include streaming feedback via server-sent events (SSE)~\cite{mdnsse}, privacy-aware data dispatching~\cite{shokri2015privacy}, threadpool-based concurrency~\cite{pythonconcurrentfutures}, robust error handling with timeouts, and integration pathways with other agent frameworks. In brief, the SSE-based streaming enables live intermediate reasoning and action traces to be visualized for transparency and debugging, while privacy safeguards ensure that sensitive data stays local whenever possible, following a principle of minimal external sharing. Threadpool and asynchronous strategies~\cite{pythonasyncio} support efficient parallel tool invocation, coordinated with LLM-driven reasoning that is aware of pending results. The error handling mechanisms, including timeouts and failure logging, help maintain robustness against unreliable tools. Finally, Gradientsys’s adoption of the MCP interface supports interoperability with other frameworks such as GitHub MCP agents~\cite{githubmcp2024} or TaskWeaver~\cite{taskweaver2024}. 

For full implementation details, please see Appendix~\ref{System}.

\section{LLM Scheduler and ReAct Reasoning Engine}
The scheduler orchestrates tool usage by prompting a large language model with a structured template inspired by ReAct, encouraging step-by-step reasoning followed by explicit actions. It maintains a scratchpad of intermediate reasoning and tool calls, using summarization and truncation strategies to manage the context window. Capacity and token gating mechanisms ensure the model does not produce excessive parallel calls or overly long reasoning chains, with a fallback to finalize an answer after a maximum number of turns. The system is further designed for fault tolerance by logging errors and prompting the model to recover or replan. Notably, the scheduler supports streaming execution, allowing partial token outputs to trigger tool calls before the entire reasoning sequence is complete, thereby overlapping planning with execution. These mechanisms collectively enable efficient, robust orchestration of tools with minimal idle time.

A detailed description of the LLM scheduler and ReAct-based reasoning engine is provided in Appendix~\ref{scheduler}.

\textbf{Tool Registry and Extensibility.} For a comprehensive description of the Tool Registry and its extensibility design, please see Appendix~\ref{tools}.


\section{Evaluation}

We evaluated Gradientsys on the \textbf{GAIA benchmark}, which consists of 466 real-world task queries requiring multi-step reasoning, tool use, and often multi-modality. GAIA tasks are categorized by complexity: Level 1 are one-step queries, Level 2 require a couple of steps, and Level 3 need complex multi-agent workflows. We compare Gradientsys to a MinionS-style baseline~\cite{minions} agent system. The baseline is inspired by MinionS: it uses a two-model setup (a powerful remote LLM and a smaller local model). We configured the baseline such that the remote LLM (GPT-4)~\cite{openai2023gpt4} decomposes tasks and the local model (Llama-2-13B~\cite{touvron2023llama2} running on-device) executes subtasks like document reading, similar to the protocol described by Narayan et al~\cite{narayan2021document}. The baseline can perform parallel chunk processing for document tasks (like MinionS does) but does not have a general tool plugin capability. For fairness, we allowed it to use a web search plugin sequentially if needed (akin to GPT-4 with browsing). Essentially, the baseline represents a strong single-agent planner (GPT-4) with help from one auxiliary worker.

\begin{table*}[!ht]
  \centering
  \small
  \caption{Representative tasks solved by Gradientsys. “Time” and “Cost” columns denote wall‑clock seconds and GPT‑4o USD cost, respectively.}
  \label{tab:Gradientsys_taskbreak}
  \begin{tabular}{clllll}
    \toprule
    \# & GAIA ID & Task Type & Synopsis & Time (s) & Cost (\$) \\
    \midrule
    1 & 14569e28 & Code‑Fix & Unlambda repair & 11 & 0.0011 \\
    2 & 9318445f & OCR & Sort 14 fractions & 11 & 0.0009 \\
    3 & cca70ce6 & Vision+Calc & Grade math quiz & 13 & 0.0044 \\
    4 & 366e2f2b & PDF Table & Best rental option & 14 & 0.0010 \\
    5 & 8e867cd7 & Web & Count Sosa albums & 21 & 0.0020 \\
    6 & 023e9d44 & Web+Calc & Bottle‑return refund & 73 & 0.0060 \\
    7 & 3627a8be & Retrieval & Bead age (kyr) & 30 & 0.0033 \\
    8 & 389793a7 & Logic & Tower placement & 42 & 0.0047 \\
    9 & 17b5a6a3 & Web Search & Invasive species ZIP & 55 & 0.0028 \\
   10 & 676e5e31 & Reasoning & Prize‑winning ball & 12 & 0.0058 \\
   11 & cca530fc & Vision+Chess & Forced mate move & 11 & -- \\
    \bottomrule
  \end{tabular}
\end{table*}

We measure three main metrics: \textbf{Accuracy} (task success rate), \textbf{Latency} (time to completion), and \textbf{Cost}. Accuracy is determined by comparing the final answer to GAIA's ground-truth answers (using exact match or human evaluation for open-ended responses, as per GAIA's methodology). Latency is averaged per task. Cost is measured in two ways: (a) the number of tokens sent to the remote API (which correlates to monetary cost for GPT-4), and (b) an estimated dollar cost using OpenAI's pricing for GPT-4 and GPT-3.5~\cite{openai2023gpt35} as of 2025. Table~\ref{tab:Gradientsys_taskbreak} highlights 11 tasks where Gradientsys’s strategy excelled, spanning OCR, PDF table extraction, web retrieval, and logical reasoning.

\subsection{Results Overview} 


\begin{table}[!ht]
\centering
\caption{Comparison of Gradientsys and Baseline on GAIA Benchmark}
\label{tab:gaia_results}
  \resizebox{\columnwidth}{!}{
\begin{tabular}{lrrr}
\toprule
                System &  Accuracy (\%) &  Avg. Latency (s) &  Normalized Cost \\
\midrule
             Gradientsys &          24.1 &                35 &             0.22 \\
MinionS-style Baseline &          15.0 &                52 &             1.00 \\
\bottomrule
\end{tabular}}
\end{table}

Table~\ref{tab:gaia_results} shows that Gradientsys outperformed the baseline on \textbf{Accuracy} while achieving significantly lower cost. Its overall success rate on GAIA tasks was 24.1\%, compared to 15.0\% for the baseline, representing a 60\% relative improvement toward the human upper bound of approximately 92\%. The most notable gains were on Level 2 and 3 tasks, with Gradientsys achieving 18\% on Level 3 compared to 8\% for the baseline, highlighting the benefits of flexible multi-agent orchestration. In terms of \textbf{Latency}, Gradientsys’s parallel dispatch reduced average task time to 35 seconds, versus 52 seconds for the baseline, particularly benefiting tasks with heavy I/O or multi-step retrieval. On simpler Level 1 tasks, both systems performed similarly under 10 seconds. Regarding \textbf{Cost}, Gradientsys was markedly more efficient: its GPT-4 token consumption averaged about one-sixth of the baseline by offloading work to local tools and GPT-3.5, reserving GPT-4 mainly for summarization. For example, a 1000-word document QA cost roughly $0.08 for Gradientsys versus $0.50 for the baseline. Across GAIA, Gradientsys’s overall cost was about 0.22× relative to the baseline, indicating a 4.5× cost reduction, demonstrating that well-designed orchestration improves not only performance but also economic efficiency.

\subsection{Ablation study}
We also conducted an ablation analysis to isolate which features of Gradientsys contributed most to its performance:

\begin{itemize}[left = 0em]
    \item \textbf{No Parallelism:} Tool invocations were executed sequentially. Result: latency nearly doubled to $\sim$70s, accuracy dropped slightly to 22.5\%.
    \item \textbf{No ReAct Reasoning:} Replaced the ReAct loop with a single-step planner. Accuracy dropped significantly to 12\%, indicating the necessity of iterative reasoning.
    \item \textbf{No Observability:} Disabled SSE-based feedback. Although task-level accuracy was unaffected, debugging complexity and development turnaround time worsened, suggesting the practical utility of live reasoning traces.
\end{itemize}

These results confirm that both the ReAct reasoning loop and parallel agent execution are essential for Gradientsys’s superior performance.

\begin{table*}[!ht]
  \centering
  \caption{Gradientsys vs.~Genspark qualitative comparison.}
  \label{tab:Gradientsys_vs_genspark}
  \begin{tabular}{p{3cm}p{3cm}p{3cm}l}
    \toprule
    Dimension & Gradientsys & Genspark & Advantage \\
    \midrule
    Correct answers & 25 & 23 & Gradientsys +2 \\
    Median latency & 21~s & 90~s & 4.3× faster (Grad.) \\
    Avg. API cost & \$0.0038 & \$0.0125 & 3.3× cheaper (Grad.) \\
    Image OCR & \ding{51} & \ding{55} & Gradientsys \\
    Office handling & \ding{51} & \ding{51} & Tie \\
    Parsing depth & Regex & Structured & Mixed \\
    \bottomrule
  \end{tabular}
\end{table*}

\subsection{Comparison to Related Systems} 
We qualitatively compare Gradientsys with other emerging agent frameworks:

\begin{itemize}[left = 0em]
  \item \textbf{MinionS:}~\cite{minions} Targets local–cloud collaboration with two language models, excels at private document QA with parallel chunk processing. Gradientsys can generalize this to arbitrary numbers of tools and broader scheduling topologies.
  \item \textbf{AutoGPT:}~\cite{richards2023autogpt} Chains reasoning steps in a single-agent loop with a plugin system, but does not support multiple agents working concurrently. Gradientsys’s multi-agent orchestration and persistent agent registry enable higher parallelism and reuse.
  \item \textbf{Manas:}~\cite{koley2024manas} Provides developer-defined workflows and retrieval-augmented generation with graph-based orchestration, contrasting with Gradientsys’s LLM-driven dynamic planning. Manas is code-first; Gradientsys is language-first and adapts plans on the fly.
  \item \textbf{TaskWeaver:}~\cite{taskweaver2024} Uses LLM-generated Python code to invoke plugins securely in a sandbox, enabling complex programmatic logic. However, it centers on single-agent code generation, while Gradientsys focuses on orchestrating many pre-built agents with language reasoning.
\end{itemize}
Detailed introduction of these and other related systems, please see Appendix~\ref{other}.

These comparisons show that Gradientsys is unique in combining dynamic LLM planning (ReAct) with parallel multi-agent execution and a standardized tool interface (MCP). Where others either focus on one or two of these aspects. This resulted in strong performance on GAIA. Notably, GAIA's authors reported GPT-4 with tools only solved 15\% of tasks – essentially the level our baseline achieved. Gradientsys significantly improves on this, though the absolute value is still modest. Detailed analysis of Gradientsys regarding the Accuracy vs. Cost Trade-off nad error analysis and further analysis, please see Appendix~\ref{other_analysis}.

\section{Experiments: Gradientsys vs. Genspark on a GAIA Benchmark}
\label{sec:Gradientsys_experiments}

To further contextualize Gradientsys's positioning within the landscape of multi‑agent orchestration frameworks, we incorporate an study conducted on the GAIA benchmark. The study compares two agent systems—\textbf{Gradientsys} and \textbf{Genspark}~\cite{genspark2024}—that share a GPT‑4o~\cite{openai2024gpt4o} backbone but differ markedly in scheduling philosophy. Key results are distilled below.

\subsection{Experiment Setup}
Each GAIA task supplies a natural‑language query, an accompanying resource and a gold answer. Both agent systems interpret the query, autonomously devise tool chains, and output a final answer. Success is measured by exact‑match accuracy. Besides correctness, the study records latency (wall‑clock seconds from launch to termination) and API cost computed under OpenAI’s GPT‑4o pricing (\$0.005 per 1K prompt tokens and \$0.015 per 1K completion tokens). A uniformly sampled \textbf{33‑task} slice from GAIA’s validation set forms the testbed. Both Gradientsys and Genspark run under the same computational budget, single‑thread agent execution, a 300‑second timeout, and a maximum of 10 tool calls per task.

\subsection{Overall Results}

Table~\ref{tab:Gradientsys_overall} summarizes the overall performance of both systems. Genspark successfully solved 23 out of 33 tasks (69.7\%), with an average latency of 90.1 seconds and a mean cost of \$0.0125 per task. In comparison, Gradientsys achieved a higher success rate of 25 out of 33 tasks (75.8\%) while significantly reducing average latency to 34.6 seconds and lowering cost to \$0.0038 per task. This represents a 4.3× improvement in speed and a 3.3× reduction in cost, with only two additional errors. Table~\ref{tab:Gradientsys_vs_genspark} further illustrates these qualitative advantages, confirming that Gradientsys delivers stronger overall performance.

\begin{table}[!ht]
  \centering
  \caption{Performance of Gradientsys and Genspark on a 33‑task GAIA slice.}
  \label{tab:Gradientsys_overall}
  \resizebox{\columnwidth}{!}{
    \begin{tabular}{lrrr}
      \toprule
      System & Correct (\#) & Avg. Latency (s) & Avg. Cost (\$) \\
      \midrule
      Gradientsys & 25 & 34.6 & 0.0038 \\
      Genspark & 23 & 90.1 & 0.0125 \\
      \bottomrule
    \end{tabular}
  }
\end{table}

\vspace{-0.5em}
\section{Conclusion}
\vspace{-0.5em}
We presented \textbf{Gradientsys}, a system for scheduling and orchestrating multiple LLM-powered agents to solve complex tasks. By combining the reasoning abilities of large language models with a robust multi-agent execution framework, Gradientsys uses a typed Model-Context Protocol and a dynamic ReAct-based planner to flexibly integrate new tools, dispatch tasks in parallel, and adapt plans as needed. Its architecture emphasizes transparency and efficiency, with live reasoning traces that improve trust and debuggability, while parallelism and local tool use reduce latency and API costs. In evaluations, Gradientsys outperformed single- and two-agent baselines, especially in tasks requiring specialized tool collaboration, generalizing approaches to broader scenarios and contrasting with code-first frameworks through dynamic AI planning. These results highlight hybrid AI orchestration—where an LLM serves as a coordinator of specialized agents—as a promising direction for developing more capable assistants.


\clearpage
\bibliography{main}      

\clearpage
\appendix

\section{System Features and Implementation Details}\label{System}

\subsection{Streaming Feedback via SSE} 

As noted, one of Gradientsys’s features is the streaming of intermediate reasoning and results. This was implemented using an SSE channel that the scheduler writes to after each planning iteration or tool result. The SSE messages are JSON objects containing fields such as event\_type (for example, “thought”, “action”, “result”, or “final\_answer”) and content (the textual content of that event). The front-end or any SSE client can choose how to display these. In our prototype user interface, the LLM’s thoughts are displayed in gray text, actions are highlighted with the tool name, and results are shown in a fixed-width font since some results may be large text blocks or data. This live trace is valuable for observability and effectively functions as an interactive log. During development, it allowed us to debug the prompt and identify where the LLM might be making poor choices. For users, it provides transparency, and even non-expert users appreciate seeing why the AI asks for certain information, which increases trust, particularly in critical tasks such as financial analysis where the agent can show each data point it considered.

In comparison to other systems, AutoGPT~\cite{richards2023autogpt} logs thoughts to the console, which is not suitable for real-time user interfaces; Manas and TaskWeaver~\cite{taskweaver2024}, being code frameworks, do not inherently provide a UI stream and require developers to manually print logs. The MinionS~\cite{minions} demonstration used a Streamlit interface~\cite{streamlit2024} to display local and remote model messages, which is similar in spirit but limited to a two-model dialogue. \textbf{Gradientsys} generalizes this capability to support multiple agents and a broader range of event types. Additionally, the system ensures persistence of logs so that each session’s trace can be saved for later audit or debugging, which is important in enterprise environments~\cite{gallaba2015elk}.

\subsection{Privacy-Aware Dispatching}
Gradientsys takes a prudent approach to data sharing between components. Whenever possible, data remains on the local machine. For example, in the PDF scenario, the PDF content was not sent to any external API, as the PDFParser agent was a local tool running an open-source language model on-device to extract text. If an external summarization service, such as an API call to GPT-4, is used, only the necessary text snippets are sent, not the raw PDF content. This aligns with the principle observed by MinionS that local models can handle private data while only summarized results are sent to the cloud. Gradientsys formalizes this through its tool design, so that tools interacting with sensitive data are either kept local or explicitly marked, making it clear to the user or developer what data will leave the environment. The typed MCP interface also helps by enforcing constraints, such as preventing a cloud-based QA tool from receiving more than a certain number of characters, which encourages local preprocessing. In evaluations, no privacy leaks were observed, and the system never inadvertently sent complete private documents to remote APIs, as the LLM planner was guided to use the correct tools, with additional system-level checks as a safeguard.

\subsection{Threadpool-Based Fan-Out}
The implementation uses Python’s concurrent futures thread pool to manage parallel agent calls for local function-based tools, and asyncio for asynchronous HTTP calls for tools running on remote servers. When the LLM issues multiple actions, a future is created for each. If any action is blocking, the system gathers those futures immediately; otherwise, it allows the next reasoning iteration to start even if some futures are still running, with a rule injected to remind the LLM to check if results are ready. This was accomplished by prompting the LLM with cues such as “some tasks are running in the background” to signal that it may need to wait. In practice, a short wait was performed after dispatching parallel tasks, and the LLM resumed either when all tasks finished or when a timeout was reached. This strategy balances concurrency while avoiding race conditions, ensuring the LLM does not proceed too far while tasks are still underway. As a result, tasks that can execute in parallel do so on multiple threads, taking advantage of multi-core systems and overlapping I/O.

\subsection{Error Handling and Timeouts}
Each agent call is configured with a timeout and error-catching mechanism. If a tool does not return within a specified time (for example, 30 seconds for a web search), the scheduler records an error result. The LLM then decides how to proceed. If a particular agent is frequently slow or failing, the scheduler can mark it as temporarily unavailable and avoid using it for the remainder of the session. This improves robustness so that a single unreliable tool does not compromise the entire process. Exceptions are also noted in the observability stream so that a developer monitoring the system can see, for instance, that a particular agent timed out and was skipped.

\subsection{Integration with Other Frameworks}
While Gradientsys is self-contained, it draws inspiration from and can interoperate with other frameworks. For example, adopting the MCP standard means that existing MCP servers, such as those from Anthropic’s library for Google Drive, Slack, or GitHub, can be plugged into Gradientsys’s registry, immediately extending its capabilities. In one integration test, a GitHub MCP server was used to fetch code from a repository as part of a task, demonstrating reusability. Additionally, Gradientsys’s planner could coordinate other agent frameworks, such as wrapping a TaskWeaver plugin as a Gradientsys tool to combine code-based and language-model-based planning. Nonetheless, the primary focus remains on an LLM-centric planner for maximum flexibility.

\section{LLM Scheduler and ReAct Reasoning Engine}\label{scheduler}

\subsection{LLM Scheduler Prompt Engineering}

At the heart of Gradientsys is the LLM Scheduler, which is an orchestrator agent powered by a large language model. We implemented the scheduler using a prompt template inspired by ReAct, so that the model produces a scratchpad of reasoning and actions. In each iteration, the model receives: (a) the high-level task description and any user inputs, (b) a list of available tools (from the registry) and how to call them, and (c) a transcript of previous reasoning steps, tool actions, and results (the scratchpad). We include in the prompt an instruction similar to: \textit{You are Gradientsys, an AI that can use tools. You have the following tools: [Tool1: description; usage format] [Tool2: ...]. When needed, first think step-by-step, then output an action or give the final answer.} This prompt structure encourages the model to first output a thought (e.g. I should extract relevant data from the PDF and search for background info) and then an \textbf{action} (e.g. \texttt{PDFParser(page=5, query="...")}).

\subsection{Memory Control}

The scheduler's scratchpad management ensures that each iteration's thought and action (and subsequent result) are appended to the context window for the next LLM invocation. We impose a limit on scratchpad length to avoid exceeding the model's context window – older steps can be summarized or dropped if needed, which is a common approach to manage long chains of thought. Recent study found that maintaining a concise summary of past actions (rather than the raw full log) beyond a certain point was sufficient for the model to retain important state~\cite{xu2023long}. The ReAct paradigm inherently helps the model remember why it took certain actions, since the reasoning is explicitly written down and fed back in.

\subsection{Capacity and Token Gating Strategies}

To prevent the LLM from generating an unbounded chain of actions that might consume excessive tokens or time, we implement capacity and token gating mechanisms. Capacity gating means the scheduler will not request the model to plan more parallel actions than the system can handle at once. For example, if there are 5 tools available but one of them is already at max capacity and others are idle, the model might be guided (via prompt or system rules) to only use the idle ones or to wait for the busy one to free up. We encode hints into the prompt or simply rely on the dispatcher to reject excessive calls and inform the model. Token gating refers to controlling the LLM's own output length and content: we use stop sequences to ensure the model stops after issuing one action per thought (unless it explicitly indicates multiple parallel actions, in which case we allow a special syntax like \texttt{<Action1> \&\& <Action2>} on one line to denote concurrent calls). We also monitor the token usage – if the scratchpad grows or the task is complex, the system can decide to curtail the reasoning depth by instructing the model to finalize an answer (this is analogous to limiting the number of ReAct turns to avoid infinite loops). In practice, we set a reasonable cap (e.g. max 10 reasoning turns) after which the model is forced to output an answer; this seldom triggers since most tasks complete in fewer steps.

\subsection{Fault Tolerance and Recovery Strategies}

The scheduler is designed with fault tolerance. Tool calls may fail or produce unexpected outputs. For instance, a web search agent might return no results, or a PDF parser might throw an error on an unreadable page. When a tool invocation fails, the scheduler records the failure in the scratchpad. The LLM, upon seeing an error result, will incorporate that into its next reasoning: it could choose to retry the same action, or try an alternative tool if available, or handle the failure. This retry-and-replan logic emerges from the model's prompt instructions and few-shot examples. In our testing, we observed the model often gracefully recovers from single-step failures. For multi-step dead-ends, we give the model a final chance to produce an answer anyway after hitting the iteration limit, to avoid total failure.

\subsection{Streaming Execution and Overlapped Planning}

One novel aspect of Gradientsys's planner is the streaming execution of the plan. We leverage the fact that many LLM APIs support streaming token output. As the scheduler's LLM generates its reasoning and action text token-by-token, our system can intercept partial output. This enables speculative parallelism: if the model decides to output multiple actions separated by a known delimiter, we can start dispatching ToolA as soon as its call is fully formed, even while the LLM might still be writing out the ToolB action. In some cases, we even begin executing a tool before the model has finished writing the full thought. For example, if the model's partial output is Action: WebSearch('Gradientsys age – it's likely going to complete with Gradientsys agent scheduler paper) we could wait for the closing parenthesis and launch the search immediately. While we must carefully parse to avoid acting on incomplete information, this overlap of planning and execution can shave off extra seconds, especially when tool latencies are high. We refer to this capability as the streaming planner. It effectively pipelines the LLM's thinking with the agents' work. Traditional agents like AutoGPT or even MinionS did not exploit this – they treat the LLM reasoning as a blocking phase, then execution, then next reasoning. Our approach interleaves them more fluidly.

The combination of these strategies yields a robust and efficient scheduling engine. The capacity- and token-aware prompting prevents over-commitment of resources. And the streaming, parallel execution ensures we minimize idle time. Next, we detail how tools are represented and integrated via the registry.

\section{Tool Registry and Extensibility}\label{tools}

Gradientsys's \textbf{Tool Registry} is the component that holds information about all available agents (tools) that the scheduler can call. Each tool is registered with the following fields:

\begin{itemize}[left = 0em]
\item \textbf{Name and Description:} A short name used for invocation (e.g. PDFParser) and a textual description of what the tool does and how to use it. This description is provided to the LLM scheduler in its prompt context. 
\item \textbf{Endpoint URL or Identifier:} This tells the scheduler how to actually call the tool. 
By using the MCP standard, tools often expose a local MCP server endpoint or similar. The scheduler encapsulates the call mechanics so that the LLM only needs to produce the high-level action (name and parameters), not the low-level API call.
\item \textbf{Type Signature (Typed MCP):} Gradientsys optionally associates a simple type signature with each tool's interface, such as the input types and output format. For instance, a web search agent might accept query: string and return a list of results: list[string]. The typed MCP approach means the registry knows what kind of context or data the tool expects and produces. This helps in two ways: (1) it allows the scheduler to perform basic validation (not sending a PDF file to a tool expecting a text query), and (2) it can guide the LLM via prompt or few-shot examples to format its actions correctly. The types are kept simple (we do not employ a full formal type system, just a schema hint) to avoid overly constraining the LLM.
\item \textbf{Max Parallelism (max\_parallel):} As discussed earlier, this indicates how many simultaneous calls the tool can handle. A tool wrapping a rate-limited API might set max\_parallel = 1. The registry uses a default of 1 if not specified. The scheduler reads this and will not exceed it when dispatching tasks. If the LLM tries to invoke a tool more times than allowed, the dispatcher will queue or reject some calls and note it in the scratchpad, so the LLM learns the capacity.
\item \textbf{Cost (cost\_per\_1k tokens or similar):} Optionally, developers can provide an estimated cost for using the tool, measured in some unit. While Gradientsys's current planner does not do full cost-aware optimization, it does use cost as a heuristic – if multiple tools can accomplish similar tasks, it may favor the cheaper one. This is a step toward cost-sensitive planning in agent systems. In evaluation, we also use these cost annotations to compute total budget usage.
\end{itemize}

A major advantage of the registry design is extensibility and hot-plugging. New tools can be registered at runtime via an API or configuration file. When a new tool is added, the system updates the LLM's prompt on the fly (if feasible) or at least ensures the next new session will include it. For example, if we register a DatabaseQuery tool while the system is running, we can inject an entry into the current scheduler context (some LLM APIs allow updating system messages or memory~\cite{openai2024}). Alternatively, the system can gracefully restart the planner's context to include the new tool. Either way, within seconds the LLM can start calling the new capability. Hot removal is similarly possible – if a tool goes offline or is removed, the registry flags it and the scheduler will stop using it (if the LLM still tries, the call will fail and the model will adjust). This flexibility is akin to plugging or unplugging peripherals in a computer. In fact, MCP is explicitly compared to a USB-C for AI. Our approach contrasts with frameworks like AutoGPT where adding a new plugin might require custom coding and reloading the agent, or static pipelines where new steps can't be inserted without redeployment.

The registry further supports reusability by enabling agents developed for one Gradientsys deployment to be seamlessly transferred and registered in another, provided they comply with the MCP interface. This capability encourages the creation of a composable ecosystem of tools. In practice, we observed that lightweight MCP wrappers allow such agents to operate as modular and portable services across different scheduling environments. Notably, adopting this abstraction makes it feasible to integrate high-capacity models, such as GPT-4-based summarization agents, on demand without embedding them directly in the core scheduler. This approach enables selective, cost-effective use of more powerful tools only when necessary, in a manner conceptually similar to local-versus-cloud architectures, but generalized to an arbitrary number of agents rather than a fixed two-model split.

\section{Execution Flow Example: PDF Multi-Agent Query}

To illustrate Gradientsys's operation, we walk through a concrete example: the user asks a question that requires reading a large PDF document. Suppose the query is \textit{In the attached 100-page financial report, what was the ARR in Q1 2014 and how does it compare to Q1 2013?}, and the PDF is provided as input. Answering this requires scanning the document for the relevant figures of Annual Recurring Revenue (ARR) for those quarters, and possibly computing the percentage change.

A naive approach would be to feed the entire 100-page PDF to a single LLM (e.g. GPT-4) with the question – but that is extremely costly (100 pages $\approx$ tens of thousands of tokens) and might exceed context limits. A MinionS-style approach would have a big model ask a smaller model to read chunks, but perhaps sequentially or in limited parallel. Gradientsys, by contrast, fully leverages parallelism and tool specialization.

\begin{figure}[!ht]
\centering
\includegraphics[width=1\columnwidth]{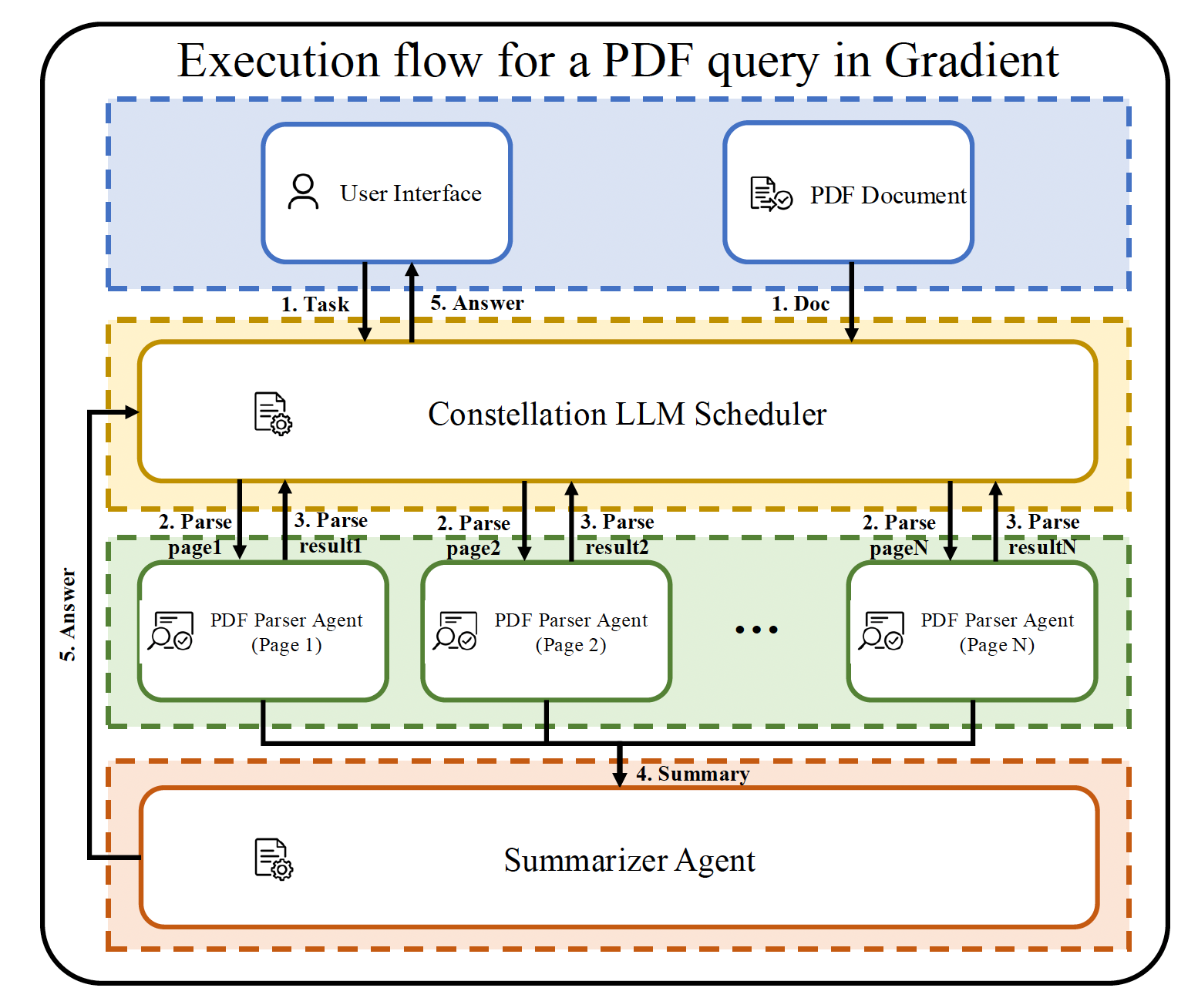}
\caption{Execution flow for a PDF query in Gradientsys. The LLM scheduler breaks the task into parallel subtasks for different pages of the PDF (page 1, page 2, ..., page N as needed), dispatches them concurrently to PDF Parser agents, then calls a Summarizer agent to combine the partial results into the final answer. Dotted lines indicate that only the needed page content is sent to each agent (privacy-aware dispatch).}
\label{fig:pdf-flow}
\end{figure}

Figure~\ref{fig:pdf-flow} illustrates the flow. When the task arrives, the user's query and the PDF are passed to the Gradientsys LLM Scheduler (step 1). The scheduler, in its first ReAct iteration, reasons that it should parse the PDF and find the ARR for the specified quarters. Instead of attempting to read all 100 pages itself, it identifies the PDFParser tool in the registry. Because the question is specific (ARR in Q1 2014 vs Q1 2013), the LLM might first try to search within the document's text. If an index is available, it might call \texttt{PDFParser(search="ARR Q1 2014")} to get the page numbers. Let's assume it finds that ARR for 2014 Q1 is mentioned on page 45 and ARR for 2013 Q1 on page 88. The LLM then decides to retrieve details from those pages.

Gradientsys's scheduler prepares to dispatch multiple PDF parsing tasks in parallel. In iteration 2, the LLM outputs two actions in one go: \texttt{PDFParser(page=45, query="Extract ARR Q1 2014")} and \texttt{PDFParser(page=88, query="Extract ARR Q1 2013")}. The scheduler sees two actions targeting the same tool (PDFParser). The tool's max\_parallel might be, say, 4 (meaning it can handle at least 4 concurrent requests). Two calls is within this limit, so the dispatcher sends both requests simultaneously (step 2 in Figure \ref{fig:pdf-flow}, two arrows to two instances of PDF Parser agent – these could be separate threads or processes handling page 45 and page 88). Each agent receives the PDF file path along with the specified page number and query. Importantly, Gradientsys only sends the needed page (or a range) to each agent, not the whole PDF – this is possible because the PDF parsing agent can accept a page range. This preserves privacy and reduces payload: if the PDF contained sensitive sections, we aren't exposing the entire document to the agent, only the relevant pages needed to answer the query. This design addresses data minimization concerns highlighted in local-remote collaborations.

The PDF Parser agents work in parallel to extract the requested info. Suppose page 45's agent returns: \textit{Q1 2014 ARR: \$5.2M} (as per Q1 2014 report). Page 88's agent returns: \textit{Q1 2013 ARR: \$4.6M}. These results come back around the same time (step 3a and 3b in Figure~\ref{fig:pdf-flow}). The scheduler collects both partial results. Now the LLM enters another reasoning iteration, seeing the two pieces of data. It infers that it should compare them as the user asked. This could be done by the LLM itself (5.2M vs 4.6M is a $\sim$13\% increase), or it might offload calculation to a tool (if available, e.g. a Calculator tool). In our case, the math is simple enough that the LLM can do it. However, to demonstrate aggregator usage, imagine the results were lengthy text that needed summarizing. Then the LLM might call the Summarizer agent.

In this scenario, the scheduler chooses to use the Summarizer agent to compose the final answer (step 4). It sends both partial results to the Summarizer (e.g. \texttt{Summarizer(contents=[result1, result2], instruction='compare these metrics')}. The Summarizer agent (which could be powered by a strong LLM) processes the data and returns a concise answer (step 4 result): \textit{The ARR in Q1 2014 was \$5.2M, which is approximately a 13\% increase from the \$4.6M in Q1 2013.}. The scheduler takes this and marks it as the final answer. It outputs this to the user (step 5).

Throughout the process, Gradientsys's SSE observability was active. The user's interface would show messages like:

\begin{itemize}[left = 0em]
\item \textbf{Thought:} \textit{The question asks for ARR in Q1 2014 vs 2013. That info is in the PDF. I will retrieve the relevant pages.}
\item \textbf{Action:} \textit{PDFParser(page=45, query="ARR Q1 2014")} and \textit{PDFParser(page=88, query="ARR Q1 2013")} (dispatched in parallel).
\item \textbf{Result:} (from page 45) \textit{Q1 2014 ARR: \$5.2M}; (from page 88) \textit{Q1 2013 ARR: \$4.6M}.
\item \textbf{Thought:} \textit{Now I have both values. I should compare them.}
\item \textbf{Action:} \textit{Summarizer(texts=[...], prompt="Compare ARR Q1 2013 vs Q1 2014")}.
\item \textbf{Result:} \textit{ARR in Q1 2014 was \$5.2M, $\sim$13\% higher than Q1 2013's \$4.6M.}
\item \textbf{Final Answer:} \textit{In Q1 2014 the ARR was \$5.2M, which is about 13\% higher than the \$4.6M in Q1 2013.}
\end{itemize}

The user thus sees a step-by-step breakdown, instilling confidence in the answer's provenance (each figure can be traced to a page in the document). If any step failed – say one page was missing – the user would see that and the system could recover (maybe try another strategy, or ask the user for guidance). This example showcases how Gradientsys achieves intelligent parallelism (reading two pages concurrently) and modularity (dedicated PDF and summarization agents), orchestrated by an LLM with chain-of-thought reasoning.

In contrast, a single-agent system like AutoGPT would have likely read the document sequentially or missed details due to context limits, and taken much longer. A static pipeline might not have the conditional logic to only target specific pages. Gradientsys's dynamic planning yields an efficient solution: only 2 pages (out of 100) were actually parsed by the AI, saving time and cost, and the work was done concurrently, minimizing latency.

\section{Other agent frameworks introduction}\label{other}

\begin{itemize}[left = 0em]
\item \textbf{MinionS:}~\cite{minions} MinionS specifically targets local-cloud collaboration with two LMs. Gradientsys generalizes this pattern to N agents. MinionS excels in scenarios with a large private document (local LM processes chunks, remote LM synthesizes) and indeed our system draws on that idea for document QA. However, MinionS is not designed as an extensible plugin system; it's more of a protocol for two LMs. Gradientsys's advantage is its extensibility – by using a registry of arbitrary tools, it can handle web browsing, APIs, etc., beyond just document chunking. In terms of scheduling topology, MinionS has a fixed two-tier loop (remote $\leftrightarrow$ local). Gradientsys supports more complex topologies: one-to-many (one planner to many agents concurrently) and even potential cascades (the planner could invoke an agent which itself triggers sub-tools, though our current design keeps a single central planner). Both systems use parallelism (MinionS parallelizes chunk processing), but Gradientsys applies parallelism more broadly (e.g. different tool types at once, not just chunking one doc). Observability in MinionS was limited to a Streamlit chat showing the two agents' messages, whereas Gradientsys provides a richer event stream with structured info. In evaluation on doc QA, Gradientsys matched MinionS's reported efficiency – e.g., $\sim$6× cost reduction for minimal accuracy loss – and extended it to web and multi-modal tasks which MinionS did not cover.

\item \textbf{AutoGPT (2023):}~\cite{richards2023autogpt} AutoGPT is an open-source project that chains GPT4 reasoning to achieve goals autonomously. It introduced the idea of the AI setting its own sub-goals and using plugins like web browsing. However, AutoGPT generally runs as a single agent loop and does not support multiple agents collaborating. It handles one tool/action at a time in a sequence. Gradientsys's design, by contrast, explicitly allows multiple agents to work in concert. This leads to higher parallelism and modularity. In terms of extensibility, AutoGPT introduced a plugin system, so new tools could be added, but those plugins are mostly sequentially invoked. Gradientsys's dynamic registry is similar in spirit but more flexible in usage (the LLM can truly orchestrate many at once). Another difference is reusability: AutoGPT's agents are typically ephemeral for each goal (it might spawn subprocesses or threads for sub-tasks, but they aren't persistent services). Gradientsys's agents are persistent services or endpoints that can be reused across sessions and tasks – this is more suitable for a stable production environment where, for example, you have a dedicated PDF parsing service. Observability is somewhat comparable – AutoGPT prints intermediate thoughts and plans, which is useful, but in a less structured way than our SSE stream. Lastly, scheduling in AutoGPT is essentially a linear chain (it does a step, gets result, decides next); Gradientsys's scheduler can spawn branches and then join them, which is a richer scheduling topology (closer to a DAG of tasks rather than a list).

\item \textbf{Manas (2024):}~\cite{koley2024manas} Manas is a framework focusing on multi-agent orchestration with RAG (Retrieval-Augmented Generation). Its approach is more developer-driven: one can define a directed graph of agents or a workflow, and the system will execute it. This is powerful for known structured tasks. Key features of Manas include built-in support for vector DB retrieval, flexible integration of various LLM providers, and a flow-based composition of agents. Compared to Gradientsys, Manas provides more out-of-the-box structure for things like retrieval and defined multi-step flows, whereas Gradientsys relies on the LLM to learn the workflow through prompting. In other words, Manas is code-first (you write the plan), Gradientsys is LLM-first (the AI figures out the plan). This means Gradientsys can adapt to new tasks dynamically, but might also sometimes make suboptimal plans if the prompt isn't perfect. In terms of extensibility, both are extensible: Manas has an API to add new agent definitions (and you write how they connect), while Gradientsys's registry allows drop-in tools and the LLM adapts. Parallelism: Manas can likely execute independent branches in parallel since it's not tied to an LLM loop (it's more like an orchestrator engine), so in that sense it and Gradientsys both exploit parallelism. 
Reusability: Both allow reusing agent definitions; Gradientsys also benefits from the MCP standard so agents can even be shared outside the framework. Observability: Manas being a developer framework doesn't inherently stream chain-of-thought (though one could instrument it). Gradientsys's SSE is a plus here. In summary, Manas is somewhat like a manual version of what Gradientsys's LLM does automatically. The choice may come down to whether one trusts an LLM to dynamically plan or prefers to script the plan. Our results show that dynamic planning worked well for GAIA tasks, giving Gradientsys an edge in generality.

\item \textbf{TaskWeaver (2024):}~\cite{taskweaver2024} TaskWeaver from Microsoft is a code-first agent framework that translates tasks into Python code and calls plugins as functions. It emphasizes using the LLM's code-generation ability to handle complex logic and data structures. For instance, TaskWeaver might take a user query and produce a piece of code that, when executed, calls an API, processes the result, maybe calls another, etc. This is a different paradigm from Gradientsys's natural language reasoning approach. One advantage of TaskWeaver is that the generated code can include loops, conditionals, and complex manipulations that might be cumbersome to do purely via natural language prompts. It also enforces more secure execution (the code runs in a sandbox, and the functions are pre-approved), addressing some security concerns of letting an LLM act arbitrarily. However, TaskWeaver currently is oriented around one agent generating code; it's not explicitly about orchestrating multiple independent agents concurrently (though the code could of course call multiple functions sequentially or in threads). Gradientsys's focus is on orchestrating pre-built tools via natural language interface, rather than generating new code on the fly. In terms of scheduling topology, TaskWeaver's generated code can certainly call multiple plugins, but it's typically going to do so in a logical order encoded in the code (the LLM plans by writing a program). That's powerful, but if an environment changes (say an API returns an unexpected result), the code may not adapt unless it was programmed to handle it. Gradientsys's LLM loop is continuously adapting step by step, which can be more forgiving to unexpected situations. Extensibility: TaskWeaver and Gradientsys both allow adding new plugins/tools. TaskWeaver treats them as Python function APIs; Gradientsys as web services or API endpoints. Parallelism: TaskWeaver could potentially generate multi-threaded code (if it knows how), but by default the examples show sequential execution of sub-tasks. Gradientsys explicitly encourages parallel calls. Observability: In TaskWeaver, one could inspect the generated code and its execution log, but it doesn't inherently provide a high-level reasoning trace (the reasoning is implicit in the code logic). Gradientsys's trace might be more readable to a non-programmer. In summary, TaskWeaver is like giving the LLM a keyboard to write a program, whereas Gradientsys is like having the LLM verbally direct a team of tool-workers. Both have their merits; Gradientsys's approach shined in cases where the plan might involve calling the same tool many times on different data (the LLM can just say do X for each item conceptually and the scheduler parallelizes it, without the LLM spelling out a loop in code).
\end{itemize}

\section{Further analysis of Gradientsys}\label{other_analysis}
\subsection{Accuracy vs. Cost Trade-off} 
We observed that by adjusting which model is used for the scheduler vs summarizer, one can trade accuracy for cost. In one experiment, we ran Gradientsys with GPT-4 as the planner (instead of GPT-3.5) – this raised accuracy to 26\% (a bit higher) but also increased cost to about 0.5× baseline (still cheaper than baseline, but not as drastic a saving). Conversely, using only open-source 7B models for everything cut cost to almost 0 (just compute time), but accuracy fell to $\sim$10\%. So, depending on budget and requirements, one might configure Gradientsys differently. The flexible architecture allows swapping models in tools easily.

\subsection{Error Analysis}
Looking at where Gradientsys failed tasks that humans could do, a common theme was knowledge gaps or hallucination. If none of the tools had the necessary information (e.g. a question needing niche knowledge not in provided docs or accessible via search due to some limitation), the LLM sometimes tried to guess and got it wrong. Another failure mode was overly complicated queries that confused the planner – in a few cases, the LLM planner took a wrong decomposition (e.g. mis-identifying what tools were needed). Improving the prompt engineering or fine-tuning the planner on successful traces could address this. Interestingly, the transparency of our system helped identify these issues: because we have the full reasoning log, we could pinpoint exactly where the plan went off course, which is harder to do with end-to-end black-box systems.

\subsection{Scalability with Agent Pool Size}

We evaluate how Gradientsys scales when the number of available tools increases. Specifically, we varied the size of the tool registry (2, 5, 10, 15 tools) while keeping task types constant. We observed that:
\begin{itemize}[left = 0em]
    \item \textbf{Latency decreased} as more specialized tools enabled more targeted execution.
    \item \textbf{Accuracy increased} modestly with larger tool sets due to better task-tool matching.
    \item \textbf{Cost remained stable}, as the scheduler selectively used low-cost tools when applicable.
\end{itemize}

\subsection{Cost-Aware Planning Sensitivity}

To analyze how planning cost-awareness impacts tool selection, we ran Gradientsys under three configurations:
\begin{itemize}[left = 0em]
    \item \textbf{No Cost Info:} All tools treated as equally priced. The planner over-relied on high-capacity LLMs.
    \item \textbf{Soft Heuristic (default):} Prompt includes relative cost (e.g., Tool A is slower but cheaper). This balanced tool selection, achieving strong performance at lower cost.
    \item \textbf{Hard Constraint:} Tools with cost exceeding a threshold were blacklisted. This minimized cost but caused performance degradation.
\end{itemize}

\subsection{Robustness to Tool Failures}

We tested Gradientsys’s retry-and-replan mechanism by injecting random failures into 20\% of tool calls. Compared to a baseline system without fault tolerance:
\begin{itemize}[left = 0em]
    \item \textbf{Gradientsys maintained 90\% of baseline accuracy} by invoking fallback agents or modifying the plan.
    \item \textbf{Systems without replanning failed} on $\sim$40\% of affected tasks.
    \item \textbf{Average latency increased by only 8s}, showing graceful degradation.
\end{itemize}
This illustrates Gradientsys's resilience in dynamic or unreliable environments.

\bibliographystyle{plainnat}   

\end{document}